\documentstyle[12pt]{article}

\font\twlgot =eufm10 scaled \magstep1
\font\egtgot =eufm8
\font\sevgot =eufm7
\font\twlmsb =msbm10 scaled \magstep1
\font\egtmsb =msbm8
\font\sevmsb =msbm7

\newfam\gotfam

\textfont\gotfam\twlgot
\scriptfont\gotfam\egtgot
\scriptscriptfont\gotfam\sevgot

\newfam\msbfam
\textfont\msbfam\twlmsb
\scriptfont\msbfam\egtmsb
\scriptscriptfont\msbfam\sevmsb
\def\Bbb{\protect\pBbb}
\def\pBbb{\relax\ifmmode\expandafter\Bb\else\typeout{You cann't use
Bbb in text mode}\fi}
\def\Bb #1{{\fam\msbfam\relax#1}}

\def\thebibliography#1{\section*{
References}\list
  {\arabic{enumi}.}{\settowidth\labelwidth{#1}\leftmargin\labelwidth
    \advance\leftmargin\labelsep
    \usecounter{enumi}}
    \def\newblock{\hskip .11em plus .33em minus .07em}
    \sloppy\clubpenalty4000\widowpenalty4000
    \sfcode`\.=1000\relax}

\def\op#1{\mathop{\fam0 #1}\limits}

\newcommand{\beq}{\begin{equation}}
\newcommand{\eeq}{\end{equation}}
\newcommand{\ben}{\begin{eqnarray}}
\newcommand{\een}{\end{eqnarray}}
\newcommand{\be}{\begin{eqnarray*}}
\newcommand{\ee}{\end{eqnarray*}}
\newcommand{\bea}{\begin{eqalph}}
\newcommand{\eea}{\end{eqalph}}

\newcommand{\cL}{{\cal L}}
\newcommand{\cE}{{\cal E}}
\newcommand{\cH}{{\cal H}}

\newcommand{\vt}{\vartheta}

\newcommand{\la}{\lambda}

\newcommand{\om}{\omega}
\newcommand{\Om}{\Omega}
\newcommand{\m}{\mu}

\newcommand{\g}{\gamma}

\newcommand{\w}{\wedge}

\newcommand{\dr}{\partial}
\newcommand{\ot}{\otimes}

\newenvironment{eqalph}{\stepcounter{equation}
\setcounter{equationa}{\value{equation}}
\setcounter{equation}{0}

\begin{eqnarray}}{\end{eqnarray}
\setcounter{equation}{\value{equationa}}}

\newcommand{\mar}[1]{}

\begin{document}
\hbox{}

{\parindent=0pt

{\large \bf The bracket and the evolution operator in covariant
Hamiltonian field theory}
\bigskip

{\sc G.Sardanashvily}
\bigskip

\begin{small}

Department of Theoretical Physics, Moscow State University,
117234 Moscow, Russia

E-mail: sard@grav.phys.msu.su 

URL: http://webcenter.ru/$\sim$sardan/

\bigskip
No bracket determines the evolution operator in covariant (polysymplectic and
multisymplectic) Hamiltonian field theory.
 
\end{small}
}

\bigskip
\bigskip

The covariant Hamiltonian field theory is
mainly developed in its multisymplectic and
polysymplectic variants, related to the two different Legendre
morphisms in the first 
order calculus of variations on fibre bundles (see [1-5] for a
survey and [6] for other generalizations of symplectic formalism).

Recall that, given a fibre bundle
$Y\to X$ coordinated by $(x^\la,y^i)$, every first order Lagrangian 
\be
L=\cL\om: J^1Y\to\op\w^nT^*X, \quad \om=dx^1\w\cdots dx^n, \quad n=\dim X,
\ee
yields the Legendre map of the jet manifold $J^1Y$ to the Legendre bundle 
\mar{br12}\beq
\Pi=\op\w^nT^*X\op\ot_YV^*Y\op\ot_YTX, \label{br12}
\eeq
coordinated by $(x^\la,y^i,p^\la_i)$. The $\Pi$ is provided 
with the canonical polysymplectic form
\mar{br2}\beq
\Om_\Pi =dp_i^\la\w dy^i\w \om\ot\dr_\la, \label{br2}
\eeq
and is regarded as the polysymplectic phase space of fields.

The multisymplectic phase space of fields is 
the homogeneous Legendre bundle
\mar{br}\beq
Z= T^*Y\w(\op\w^{n-1}T^*X), \label{br}
\eeq
coordinated by $(x^\la,y^i,p^\la_i,p)$ and equipped with the 
canonical multisymplectic form
\mar{br3}\beq
\Om=d\Xi= dp\w\om + dp^\la_i\w dy^i\w\om_\la, \qquad
\om_\la=\dr_\la\rfloor \om. \label{br3}
\eeq
There is the one-dimensional affine bundle
\mar{br1}\beq
\zeta: Z\to\Pi, \label{br1}
\eeq
which is trivial if $X$ is orientable. Given a
section
$h$ of this bundle, the pull-back $h^*\Xi$ of the multisymplectic
Liouville form $\Xi$ on $Z$ is a
polysymplectic Hamiltonian form on $\Pi$. 

It is natural that one attempts to generalize a Poisson
bracket on symplectic manifolds to polysymplectic and multisymplectic
manifolds in order to obtain the covariant canonical
quantization of field theory. Different variants of such a bracket have
been suggested (see the recent works [7,8]). It however seems that
no canonical bracket corresponds to the $TX$-valued polysymplectic form
(\ref{br2}), unless dim$X=1$ [9,10]. On
the contrary, using the exterior multisymplectic form (\ref{br3}), one
can associate multivector fields to exterior forms  
on the fibre bundle $Z$ (\ref{br}) (but not to all of them), and can
introduce a desired bracket of these forms
via the Schouten--Nijenhuis bracket of multivector fields [8]. 

At the same time, I aim to show that no
bracket determines the evolution operator in 
polysymplectic and multisymplectic Hamiltonian formalism, 
including the case of dim$X=1$ of 
(time-reparametrized) time-dependent mechanics.
It should be noted that, written as a bracket, the evolution operator
is quantized, and it determines the Heisenberg equation.  

Recall the following relationship between first order
dynamic equations, connections, multivector fields and evolution
operators on a fibre bundle.

(i) Let $\pi:Q\to X$ be a fibre bundle coordinated by $(x^\m,q^a)$.
Being a section of the jet bundle $J^1Q\to Q$, any connection
\mar{br10}\beq
\g=dx^\m\ot (\dr_\m +\g^a_\m\dr_a), \label{br10}
\eeq
on $Q\to X$ defines the first order differential operator
\mar{br5}\beq
D: J^1Q\ni (x^\m,q^a,q^a_\m) \to (x^\m,q^a,
q^a_\m-\g^a_\m(x^\nu,q^b))\in T^*X\ot VQ \label{br5}
\eeq
on $Q\to X$ called the covariant differential with respect to $\g$. 
The kernel of this differential operator is a closed imbedded subbundle
\mar{br6}\beq
q^a_\m-\g^a_\m(x^\nu,q^b)=0 \label{br6}
\eeq
of $J^1Q\to X$, i.e., it is the first order dynamic
equation on a fibre bundle $Q\to X$. Conversely, any first order
dynamic equation 
on $Q\to X$ is of this type. 

(ii) Let $HQ\subset TQ$ be the horizontal
distribution defined by a connection $\g$. If $X$ is
orientable, there exists a nowhere vanishing
global section of the exterior product $\op\w^n HQ\to Q$. It is a locally
decomposable $\pi$-transverse $n$-vector field on $Q$. Conversely,
every multivector field of this type on $Q\to X$ yields a connection and,
consequently, a first order dynamic equation on this 
fibre bundle [11]. 

(iii) Given a first order dynamic equation $\g$ on a fibre bundle $Q\to X$, 
the corresponding evolution operator $d_\g$ is defined as 
the pull-back $d_\g$ onto the shell (\ref{br6}) of the horizontal differential 
\be
d_H=dx^\m(\dr_\m +q^a_\m\dr_a)
\ee
acting on smooth real functions on $Q$. 
It reads
\mar{br11}\beq
d_\g f=(\dr_\m +\g^a_\m\dr_a)f dx^\m, \qquad f\in C^\infty(Q). \label{br11}
\eeq
A glance at this expression shows that $d_\g$ is projected onto $Q$,
and it is a first order differential operator on functions on $Q$.
In particular, if a function $f$ obeys the evolution equation $d_\g
f=0$, it is constant on any solution of
the dynamic equation (\ref{br6}). 

In Hamiltonian dynamics on $Q$, a problem is to represent
the evolution operator (\ref{br11}) as a bracket of $f$ with some
exterior form on $Q$.

First of all, let us study the case of fibre bundles
$Y\to X$ over a one-dimensional orientable connected manifold $X$
(i.e., $X$ is 
either $\Bbb R$ or $S^1$). In this case, the Legendre
bundle $\Pi$ (\ref{br12}) is isomorphic to the vertical cotangent 
bundle $V^*Y$ of $Y\to X$ coordinated by $(x,y^i,p_i)$, and the
polysymplectic form $\Om_\Pi$ 
(\ref{br2}) on $V^*Y$ reads
\mar{br15}\beq
\Om_\Pi=dp_i\w dy^i\w dx\ot\dr_x. \label{br15}
\eeq
Accordingly, the homogeneous Legendre bundle (\ref{br}) is the
cotangent bundle $T^*Y$, coordinated by $(x,y^i,p,p_i)$, and the
multisymplectic form (\ref{br3}) comes to the canonical symplectic form
\mar{br16}\beq
\Om=dp\w dx + dp_i\w dy^i \label{br16}
\eeq
on $T^*Y$. Then $\zeta$ (\ref{br1}) is the canonical trivial fibre bundle
\mar{br17}\beq
\zeta: T^*Y\to V^*Y. \label{br17}
\eeq

Due to this fibration, the vertical cotangent bundle $V^*Y$ is provided with
the canonical Poisson bracket
\mar{br18}\beq
\{f,f'\}_V=\dr^if\dr_i f'-\dr_if\dr^i f', \qquad f,f'\in C^\infty(V^*Y),
\label{br18}
\eeq
given by the relation
\be
\zeta^*\{f,f'\}_V=\{\zeta^* f,\zeta^* f'\}
\ee
where $\{,\}$ is the canonical Poisson bracket on $T^*Y$ [12,13]. The
Poisson structure (\ref{br18}) however fails to determine Hamiltonian
dynamics on the fibre bundle $V^*Q\to X$ because all Hamiltonian vector
fields with respect to this structure are vertical. At the same time,
in accordance with general polysymplectic formalism [3,4], 
a section $h$, $p\circ h=-\cH$, of the fibre bundle (\ref{br17})
yields a polysymplectic Hamiltonian
form
\mar{br19}\beq
H=p_idy^i-\cH dx \label{br19}
\eeq
on $V^*Y$. The associated Hamiltonian connection on $V^*Y\to X$ with 
respect to
the polysymplectic form (\ref{br15}) is
\mar{br20}\beq
\g_H=dx\ot(\dr_x +\dr^i\cH\dr_i -\dr_i\cH\dr^i). \label{br20}
\eeq
It defines the Hamilton equation
\be
y^i_x=\dr^i\cH, \qquad p_{ix}=-\dr_i\cH
\ee
on $V^*Y$. The corresponding evolution
operator (\ref{br11}) takes the local form
\mar{br21}\beq
d_\g f=(\dr_x f+\{\cH,f\}_V)dx, \qquad f\in C^\infty(V^*Q). \label{br21}
\eeq
The bracket $\{\cH,f\}_V$ in this expression is not globally defined 
because $\cH$ 
is not a function on $V^*Q$.
Therefore, the evolution operator (\ref{br21}) does not reduce to the Poisson
bracket (\ref{br18}).

Turn now to the symplectic manifold $T^*Y$.
Let us consider the pull-back $\zeta^*H$ of the Hamiltonian form $H$
(\ref{br19}) onto $T^*Y$. Then the difference
\mar{br22}\beq
H^*=\Xi-\zeta^*H=(p+\cH)dx \label{br22}
\eeq
is a horizontal density on the fibre bundle $T^*Y\to X$. 
It is a multisymplectic Hamiltonian form. The
corresponding Hamiltonian connection 
$\g$ on $T^*Y\to X$ is given by the condition 
\mar{br23}\beq
\g(\Om)=dH^*, \label{br23}
\eeq
where the morphism
\be
\g(\Om)=dx\w [(\dr_x +\g^p\dr_p+\g^i\dr_i+\g_i\dr^i)\rfloor \Om]
\ee
is induced by an endomorphism of $T^*Y$
determined by the tangent-valued form $\g$. We obtain
\mar{br24}\beq
\g=dx\ot (\dr_x +\g^p\dr_p +\dr^i\cH\dr_i -\dr_i\cH\dr^i), \label{br24}
\eeq
where the coefficient $\g^p$ is arbitrary. It is readily observed that
this connection projects 
to the connection $\g_H$ (\ref{br20}) on $V^*Y\to X$. As a consequence,
it defines the
evolution operator whose restriction to the pull-back of functions on
$V^*Q$ is exactly the evolution operator (\ref{br21}). But now this
operator locally reduces to the Poisson bracket
\mar{br30}\beq
d_\g f=\{p+\cH,f\}dx, \qquad f\in C^\infty(V^*Y), \label{br30}
\eeq
on $T^*Y$. However, this bracket is not globally defined, too, since $p+\cH$ is
a horizontal density, but not a function on $T^*Y$.

Furthermore, following the multivector scheme in Refs. [8], let us
introduce the function 
$\cE=\rho^{-1}(p+\cH)$
on $T^*Y$ where $\rho dx$ is some nowhere vanishing
density on $X$. The Hamiltonian vector field of $\cE$ with respect to
the symplectic form $\Om$ on $T^*Y$ reads
\be
\vt_\cE=\rho^{-1}\dr_x -\dr_x\cE\dr^p + \dr^i\cE\dr_i -\dr_i\cE\dr^i.
\ee
This vector field is horizontal with respect to the connection
(\ref{br24}) where $\g_p=-\rho\dr_x\cE$, and it determines this connection
in the form
\be
\g= dx\ot (\dr_x -\rho\dr_x\cE\dr^p +\rho \dr^i\cE\dr_i -\rho\dr_i\cE\dr^i).
\ee 
Accordingly, the evolution operator (\ref{br30}) is rewritten as
\mar{br31}\beq
d_\g f=\rho \{\cE,f\}dx, \label{br31}
\eeq
The bracket $\{\cE,f\}$ is well-defined, but $d_\g$ does not equal
this bracket because of the factor $\rho$. 

The multisymplectic bracket with the horizontal density $H^*$
(\ref{br22}) also can not 
help us since there is no Hamiltonian multivector field associated to
$H^*$ relative to the symplectic form $\Om$. 

Of course, the manifolds $X=\Bbb R$ and $X=S^1$ can be equipped with 
coordinates $x$ possessing transition functions $x'=x+$const, and one
can always choose the density $\rho=1$. Then the evolution operator 
(\ref{br31}) reduces
to a Poisson bracket in full. If $X=\Bbb R$, this is the case of
time-dependent mechanics where time reparametrization is forbidden
[14,15].

Turn now to the general case of dim$\,X>1$. In the framework of
polysymplectic formalism [3,4], a polysymplectic Hamiltonian form on the
Legendre bundle $\Pi$ (\ref{br12}) reads
\mar{br40}\beq
H=p^\la_i dy^i\w \om_\la -\cH\om. \label{br40}
\eeq
The associated Hamiltonian connection 
\be
\g_H=dx^\la\ot(\dr_\la +\g^i_\la\dr_i +\g^\m_{i\la}\dr^i_\m)
\ee
fails to be uniquely determined,
but obeys the equations
\be
\g^i_\la=\dr^i_\la\cH, \qquad \g^\la_{i\la}=-\dr_i\cH.
\ee
The values of these connections assemble into a closed imbedded subbubdle 
\be
y^i_\la=\dr^i_\la\cH, \qquad p^\la_{i\la}=-\dr_i\cH
\ee
of the jet bundle $J^1\Pi\to X$ which is the 
first order polysymplectic Hamilton equation on $\Pi$.
This equation 
is not algebraically solved for the
highest order derivatives and, therefore, it is not 
a dynamic equation. As a consequence, the evolution operator depends on
the jet coordinates $p^\la_{i\m}$ and, therefore, it is not a
differential operator on functions on $\Pi$.
Clearly, no bracket on $\Pi$ can determine such an operator.

In the framework of multisymplectic formalism, the associated
multisymplectic Hamiltonian form on the homogeneous Legendre bundle $Z$
(\ref{br}) is the horizontal density 
\mar{br41}\beq
\Om-\zeta^*H =(p+\cH)\om. \label{br41}
\eeq
It defines a set of Hamiltonian connections $\g$ projected onto Hamiltonian
connections $\g_H$ on $\Pi$. These connections are also determined by
Hamiltonian multivector fields associated to a function $\cE=\rho^{-1}(p+\cH)$
where $\rho\om$ is some nowhere vanishing density on $X$. Now we are in
the case when the density $\rho=1$ need not exist. The values of
Hamiltonian connections $\g$ form a closed imbedded subbundle of the
jet bundle $J^1Z\to X$ which is a first order differential equation on $Z$,
but not a dynamic equation. As a consequence, the corresponding
evolution operator as like as its polysymplectic counterpart fails to
be a differential operator on functions on $Z$. Therefore, no bracket on
$Z$ can determine it.

Of course, one can ignore the evolution operator problem for a time, and
can try to quantize the multisymplectic bracket on $Z$ in accordance
with some modified Dirac condition. Then it easily observed that, for
instance, the corresponding quantum
algebra does not contain functions on $Z$ because their multisymplectic
bracket is not defined. Therefore, such quantization by no means is 
generalization of quantization of symplectic and Poisson manifolds. In
my opinion, its
physical perspectives do not look promising.

\end{document}